\documentclass[preprint,showpacs,preprintnumbers,amsmath,amssymb]{revtex4}
\usepackage{amsmath,amssymb,graphics,epsfig,subfigure}
\usepackage{cases}
\usepackage{color}
\usepackage[colorlinks,
            linkcolor=blue,
            anchorcolor=blue,
            citecolor=green
            ]{hyperref}
\usepackage[colorlinks,linkcolor=blue,anchorcolor=blue,citecolor=green]{hyperref}

\begin{document}
\renewcommand{\baselinestretch}{1.3}

\title{Limiting the Number of Extra Dimensions with Shortcuts}

\author{Zi-Chao Lin$^a$\footnote{linzch22@hust.edu.cn},
        Hao Yu$^b$\footnote{yuhaocd@cqu.edu.cn},
        Yungui Gong$^c$$^a$\footnote{gongyungui@nbu.edu.cn (corresponding author)}}

\affiliation{$^{a}$School of Physics, Huazhong University of Science and Technology, Wuhan, Hubei 430074, China\\
$^{b}$College of Physics, Chongqing University, Chongqing 401331, China\\
$^{c}$Department of Physics, School of Physical Science and Technology, Ningbo University, Ningbo, Zhejiang 315211, China}

\begin{abstract}
In higher-dimensional theories, a graviton propagating in the bulk can follow a shorter path, known as a shortcut, compared to a photon propagating in a four-dimensional spacetime. Thus by combining the observations of gravitational waves and their electromagnetic counterparts, one can gain insights into the structure and number of extra dimensions. In this paper, we construct a braneworld model that allows the existence of shortcuts in a $D(=4+d)$-dimensional spacetime. It has been proven that the equations for modeling brane cosmology recover the standard Friedmann equations for the late universe. We derive analytically the graviton and photon horizon radii on the brane under the low-energy limit. With the event GW170817/GRB 170817A, we find that the number of extra dimensions has an upper limit of $d\leq9$. Because of the errors in the source redshift and time delay, this upper limit can be shifted to $d\leq4$ and $d\leq12$. Although with the joint constraint on the $\text{AdS}_{D}^{~}$ radius from torsion balance measurements, theories with large $d$ are not yet ruled out, our work provides a new way to limit the number of extra dimensions.
\end{abstract}



\maketitle

\section{Introduction}

Gravitational wave (GW) observation provides an important approach to test general relativity (GR) in the strong-field regime. Since the first detection of GWs in 2015~\cite{Abbott1}, the collaborative efforts of LIGO and Virgo have led to the observation of more than 100 GW candidates originating from astrophysical compact binary coalescences, including 3 in the first observing run~\cite{Abbott1,Abbott2}, 11 in Gravitational-Wave Transient Catalog-1 (GWTC-1)~\cite{Abbott3}, 55 in GWTC-2.1~\cite{Abbott4}, and 90 in GWTC-3~\cite{Abbott5}. Among these, the event GW170817 is the earliest observed candidate attributed to the binary neutron star (BNS) coalescence~\cite{Abbott6}. Associated with this event, a gamma ray burst (named GRB 170817A) observed by $Fermi$ Gamma-ray Burst Monitor and the spectrometer on board $INTEGRAL$ Anti-Coincidence Shield~\cite{Coulter1,Pan1} is widely believed to be its electromagnetic counterpart. Confusingly, the follow-up analyses of the event GW170817/GRB 170817A revealed a time delay of $1.74\,\text{s}$ between the arrival of two signals~\cite{Abbott7}. The standard astrophysics model is unable to provide a unique explanation for the observed time delay due to the large degree of freedom involved in modeling the emissions and propagations of GWs and their electromagnetic counterparts. So far, the event GW170817/GRB 170817A has sparked extensive research on explaining the observed time delay, while it also provides a new scheme for constraining modified theories of gravity, including extra-dimensional theories~\cite{Ezquiaga:2017ekz,Sakstein:2017xjx,Belgacem:2018lbp,Hagihara:2019ihn,Odintsov:2019clh,Ghosh1,Yu1,Visinelli1,Lin1,Lin2}.

Although the generation of the target signals exhibits model dependence~\cite{Shibata1, Rezzolla1, Tsang1, Paschalidis1, Ciolfi1, Rezzolla2}, the combined observations of GWs and their electromagnetic counterparts is mainly sensitive to the propagation of the signals. In the braneworld theory, photons are typically constrained to propagate along a four-dimensional brane (4-brane), while gravitons are allowed to propagate freely in the bulk. As a result, photons and gravitons will follow different paths between two points, experiencing different numbers of spacetime dimensions. If a graviton has the speed of light, its path could be shorter than the path of the photon and thus is called as ``shortcut'' in certain studies~\cite{Chung1,Ishihara1,Caldwell1,Wang2002}. This phenomenon is then used to explain the time delay observed in the event GW170817/GRB 170817A, and to constrain the structure of extra dimensions in specific braneworld theories~\cite{Yu1,Visinelli1,Lin1,Lin2}.

In higher-dimensional theories, the number of extra dimensions is a fundamental question. As mentioned previously, gravitons can propagate in the bulk, so the leakage of gravitons into the bulk during the propagation of GWs on the brane will result in an additional loss of GW energy, which is manifested as the measured amplitudes of GWs being weaker than the predicted results in GR~\cite{Dvali1,Deffayet1}. Apparently, the number of extra dimensions plays a decisive role in the amplitude attenuation of GWs. Generally speaking, if the spacetime carries extra space, the luminosity distance of the astrophysical compact binary coalescence individually measured by GW observations can be larger than its true value. Based on it, the dimension of the spacetime in some specific models will be constrained to $D\sim4$ by comparing the source luminosity distances of the event GW170817/GRB 170817A respectively measured by GW observations and electromagnetic wave (EMW) observations~\cite{Pardo1,Abbott8}. It is worth noting that the screening mechanisms (for instance, by considering the coupling between higher-dimensional gravity and a bulk scalar field) in these models can alter the above constraint. A large screening scale, which closely matches the source luminosity distance, implies a minimal leakage of gravitons. So the freedom on scaling screening radius only places a lower boundary on the number of  spacetime dimensions as $D>4$~\cite{Pardo1,Abbott8}. It is still hard to rule out the existence of extra dimensions by the leakage of gravitons individually.

In this work, we propose an approach to constrain the number of extra dimensions with shortcut. By considering a de Sitter (dS) 4-brane embedded in a spherically symmetric $D$-dimensional spacetime, we demonstrate that the time delay observed in the joint observations of GW170817/GRB 170817A can impose a limit to the number of extra dimensions. Combined with the results presented in Refs.~\cite{Yu1,Visinelli1}, we found that it becomes an upper limit  significantly narrowing down the number of extra dimensions for the higher-dimensional theories.

The paper is arranged as follows. In Sec.~\ref{sec2}, we embed a moving 4-brane in a $D$-dimensional anti-de Sitter (AdS) spacetime and prove that the cosmology on the 4-brane can describe the normal expansion of a $\text{dS}_{4}^{~}$ universe. Then, Sec.~\ref{sec3} is dedicated to deriving analytical expressions for the photon horizon radius and gravitational horizon radius under the low-energy limit. In Sec.~\ref{sec4}, we use the time delay observed in the event GW170817/GRB 170817A to establish an upper limit to the number of extra dimensions. Finally, we present a brief conclusion in Sec.~\ref{sec5}.

\section{Cosmological Model}\label{sec2}

Let us consider a 4-brane embedded in a $D(=4+d)$-dimensional spacetime. The ordinary matter that governs the expansion of our universe is confined on it. To study the shortcut in the following context, we will study whether the so-called brane cosmology~\cite{Ida:1999ui,Binetruy:1999ut,Deffayet:2000uy,Langlois:2002bb,Brax:2003fv,Brax:2004xh} in this model can recover the standard Friedmann equations. For the sake of simplification, we ignore the backreaction of the 4-brane to the background spacetime and suppose that the underlying gravity is $D$-dimensional GR. We consider an $\text{AdS}_{D}^{~}$ metric ($\Lambda<0$) for the bulk spacetime ($M_{4}^{~}\times R\times S_{d-1}^{~}$) given by~\cite{Olasagasti:2000gx,Lin3}
\begin{equation}\label{bmetric1}	\text{d}s_{D}^{2}=-R^{2}_{~}\text{d}T^{2}_{~}+R^{2}_{~}\text{d}\Sigma_{3}^{2}+A(R)\text{d}R^{2}_{~}+B(R)\text{d}\Omega_{d-1}^{2},
\end{equation}
where
\begin{equation}
	A(R)=\frac{(d+2)(d+3)}{-2\Lambda}\frac{1}{R^{2}_{~}},~~~~B(R)=\frac{(d+2)(d+3)}{-2\Lambda}R^{2}_{~}.
\end{equation}
The line element of the extra space ($R\times S_{d-1}^{~}$) can be labeled as
\begin{equation}
	\text{d}s^{2}_{d}=A(R) \text{d}R^{2}_{~}+B(R)\text{d}\Omega_{d-1}^{2},
\end{equation}
where
\begin{equation}
	\text{d}\Omega_{d-1}^{2}
	=\tilde{g}_{mn}^{~}\text{d}y^{m}_{~}\text{d}y^{n}_{~}\nonumber\\
	=\text{d}\theta_{1}^{2}+\dots+\text{sin}^{2}_{~}\theta_{1}^{~}\dots\text{sin}^{2}_{~}\theta_{d-2}^{~}\text{d}\theta_{d-1}^{2}
\end{equation}
is a $(d-1)$-dimensional sphere of the radius $\tilde{R}=\sqrt{(d+2)(d+3)}R/\sqrt{-2\Lambda}$. The other four-dimensional submanifold ($M_{4}^{~}$) consists of one-dimensional time and a flat three-dimensional space, whose line element is written as
\begin{equation}
    \text{d}s_{4}^{2}=-R^{2}_{~}\text{d}T^{2}_{~}+R^{2}_{~}[\text{d}r^{2}_{~}+r^{2}_{~}(\text{d}\theta^{2}_{~}+\text{sin}^{2}_{~}\theta\text{d}\phi^{2}_{~})].
\end{equation}
The 4-brane can have dynamics in such spacetime. In particular, it can move in any directions in the extra space. In this paper, we are interested in the shortcut characterized by the gravitational waves which are initially emitted from the moving 4-brane and finally return to it. As was shown in our previous work~\cite{Lin2}, the 4-brane's motion along $y^{m}_{~}$ however only contributes higher-order corrections to the shortcut, which are negligible to the low-redshift gravitational wave sources. Thus it is convenient to fix the 4-brane in $y^{m}_{~}$ direction and let the 4-brane expand in $R$ direction, leaving an expression of the 4-brane's position as follows:
\begin{equation}\label{ec1}
	R=\mathcal{R}(T),~~~y^{m}_{~}=y^{m}_{0}.
\end{equation}
Using this condition, the induced metric coupling to the ordinary matter on the 4-brane then becomes
\begin{equation}\label{im1}
	\text{d}s^{2}_{4}=-\mathcal{R}^{2}_{~}\mathcal{H}^{2}_{~}\text{d}T^{2}_{~}+\mathcal{R}^{2}_{~}\text{d}\Sigma_{3}^{2},
\end{equation}
where
\begin{equation}
	\mathcal{H}^{2}_{~}=1-\frac{(d+2)(d+3)}{-2\Lambda}\frac{\dot{\mathcal{R}}^{2}_{~}}{\mathcal{R}^{4}_{~}}.
\end{equation}
Here and after, we use a dot to denote the derivative with respect to the bulk time $T$. The above metric implies that the 4-brane we consider is homogenous, isotropic, and flat, which is consistent with our real universe. We can rewrite the metric (\ref{im1}) in the form of the Friedmann-Lema\^{\i}tre-Robertson-Walker (FLRW) metric ($k=0$):
\begin{equation}\label{im2}
	\text{d}s_{4}^{2}=-\text{d}t^{2}_{~}+a(t)^{2}_{~}\text{d}\Sigma_{3}^{2},
\end{equation}
where $t$ is the cosmic time and $a(t)$ is the scale factor. The correspondence of the metrics~\eqref{im1} and~\eqref{im2} reveals
\begin{subequations}
	\begin{eqnarray}
	  \text{d}t^{2}_{~}\!&\!=\!&\!\mathcal{R}^{2}_{~}\mathcal{H}^{2}_{~}\text{d}T^{2}_{~},\label{cbr1}\\
	  a^{2}_{~}\!&\!=\!&\!\mathcal{R}^{2}_{~}.\label{sfR1}
\end{eqnarray}
\end{subequations}
These relations will later help us to give an effective description of the brane cosmology.

With the embedding condition~\eqref{ec1}, we can define the projection tensor for the 4-brane,
\begin{equation}
	h_{MN}^{~}=\gamma_{MN}^{~}-n_{M}^{~}n_{N}^{~}=g_{MN}^{~}-\delta_{~M}^{m}\delta_{~N}^{n}g_{mn}^{~}-n_{M}^{~}n_{N}^{~},
\end{equation}
where
\begin{equation} n_{M}^{~}=\bigg(\!-\sqrt{\frac{AR^{2}_{~}}{R^{2}-A\dot{\mathcal{R}}^{2}_{~}}}\dot{\mathcal{R}},0,0,0,\sqrt{\frac{AR^{2}_{~}}{R^{2}-A\dot{\mathcal{R}}^{2}_{~}}},0,\dots,0\bigg)
\end{equation}
is the unit vector normal to the 4-brane defined on the five-dimensional submanifold ($M_{4}^{~}\times R$) and $\gamma_{MN}^{~}$ is the projection tensor for the submanifold. Then the extrinsic curvature of the 4-brane for the normal vector $n_{M}^{~}$ reads
\begin{equation}
	K_{MN}^{~}=h_{~M}^{K}h_{~N}^{L}\nabla_{K}^{~}n_{L}^{~}\,.
\end{equation}
Through the field equations, we know that it is related to the energy-momentum tensor of the 4-brane:
\begin{equation}
	T_{MN}^{~}=[(\rho+p)u_{M}^{~}u_{N}^{~}+p\,h_{MN}^{~}]\delta(R-\mathcal{R})\delta^{(d-1)}_{~}(y^{m}_{~}-y^{m}_{0}),
\end{equation}
where
\begin{equation}
	u_{M}^{~}=\Bigg(\frac{1}{\sqrt{R^{2}-A\dot{\mathcal{R}}^{2}_{~}}},0,0,0,\frac{\dot{\mathcal{R}}}{\sqrt{R^{2}-A\dot{\mathcal{R}}^{2}_{~}}},0,\dots,0\Bigg)
\end{equation}
is the unit vector on the 4-brane. The parameters $\rho$ and $p$ are, respectively, the energy density and pressure of the matter on the 4-brane. Integrating the field equations over the extra ($d-1$)-dimensional space ($S_{d-1}^{~}$), the embedding of the 4-brane then requires the effective stress-energy tensor of the 4-brane,
\begin{equation}\label{eset1}
	S_{MN}^{(5)}=\int\big(T_{MN}^{~}-\frac{1}{3}h_{MN}^{~}T\big)\text{d}^{d-1}_{~}y,
\end{equation}
to source a jump in the extrinsic curvature across the 4-brane in $R$ direction. It is described by the well-known Israel joint condition~\cite{Israel1}
\begin{equation}\label{Ijc1}
	[K_{MN}^{~}]=-\frac{1}{VM_{*}^{D-2}}\int S_{MN}^{(5)}\text{d}R,
\end{equation}
where $M_{*}^{~}$ is the fundamental mass scale of the $D$-dimensional gravity and $V$ is the volume of the extra space ($S_{d-1}^{~}$). Note that with the definition~\eqref{eset1}, the effective induced stress-energy tensor $S_{MN}^{(5)}$ is singular of order one. So a nonvanishing $[K_{MN}^{~}]=K_{MN}^{+}(T,\mathcal{R}^{+}_{~})-K_{MN}^{-}(T,\mathcal{R}^{-}_{~})=-2K_{MN}^{~}(T,\mathcal{R})$ denotes a singular hypersurface of order one in the submanifold ($M_{4}^{~}\times R$).

The nonvanishing components of the condition~\eqref{Ijc1} give the dynamics of the 4-brane as
\begin{equation}\label{dob1}
	\frac{\rho}{6VM_{*}^{D-2}}\mathcal{H}
	=
	\sqrt{\frac{-2\Lambda}{(d+2)(d+3)}}\,.
\end{equation}
Using the embedding condition~\eqref{ec1} and introducing the bare cosmological constant, $\Lambda_{\text{b}}^{~}$, on the 4-brane by $\rho\rightarrow\rho+M_{\text{Pl}}^{2}\Lambda_{\text{b}}^{~}$ with $M_{\text{Pl}}^{~}$ being the Planck scale, this equation further transforms into
\begin{eqnarray}\label{fFe1}
	H^{2}_{~}
	\!&\!=\!&\!
	\bigg(\frac{M_{\text{Pl}}^{2}\Lambda_{\text{b}}^{~}}{6VM_{*}^{D-2}}\bigg)^{2}_{~}-\frac{-2\Lambda}{(d+2)(d+3)}+\frac{2M_{\text{Pl}}^{2}\Lambda_{\text{b}}^{~}}{(6VM_{*}^{D-2})^{2}_{~}}\rho
	+\bigg(\frac{\rho}{6VM_{*}^{D-2}}\bigg)^{2}_{~}\nonumber\\
	\!&\!=\!&\!
	\frac{\Lambda_{\text{eff}}^{~}}{3}+\frac{\rho}{3M_{\text{Pl}}^{2}}+\bigg(\frac{\rho}{6VM_{*}^{D-2}}\bigg)^{2}_{~},
\end{eqnarray}
where $H$ is the Hubble parameter. Obviously, Eq.~\eqref{fFe1} tells us how the 4-brane (the universe) expands for an observer on it. Up to the order $\sim\rho$, it should be consistent with the Friedmann equations for the current state of the universe. So the bare cosmological constant in Eq.~\eqref{fFe1} is set to
\begin{equation}
	\Lambda_{\text{b}}^{~}=6\bigg(\frac{VM_{*}^{D-2}}{M_{\text{Pl}}^{2}}\bigg)^{2}_{~}
\end{equation}
to recover the normal expansion of our universe. Then, the effective cosmological constant $\Lambda_{\text{eff}}^{~}$ on the 4-brane is related to the bulk and the bare cosmological constant through
\begin{equation}\label{ecm1}
	\Lambda_{\text{eff}}^{~}=3\bigg(\frac{M_{\text{Pl}}^{2}\Lambda_{\text{b}}^{~}}{6VM_{*}^{D-2}}\bigg)^{2}_{~}-\frac{-6\Lambda}{(d+2)(d+3)}.
\end{equation}
It means that the balance between the bulk cosmological constant and the bare one models the scalar curvature of the 4-brane. For the Minkowski 4-brane, the bulk cosmological constant is
\begin{equation}
	\Lambda=\Lambda_{0}^{~}=-\frac{(d+2)(d+3)}{2}\bigg(\frac{VM_{*}^{D-2}}{M_{\text{Pl}}^{2}}\bigg)^{2}_{~}.
\end{equation}
For a $\text{dS}_{4}^{~}$ brane, we have $\Lambda>\Lambda_{0}^{~}$, and for an $\text{AdS}_{4}^{~}$ brane, we have $\Lambda<\Lambda_{0}^{~}$. The value of the bulk cosmological constant cannot be completely fixed by Eq.~\eqref{ecm1}, so we can think of it as a free parameter. In Sec.~\ref{sec4}, we will show that with the current constraint on the effective cosmological constant, the right-hand side of~\eqref{ecm1} indeed can satisfy the time delay observed in the joint observations of GW170817/GRB 170817A. Thus, the construction of a $\text{dS}_{4}^{~}$ brane in the model commits with the current observations.

\section{Shortcut}\label{sec3}

In the last section, we embed our universe as a 4-brane in the $D$-dimensional bulk spacetime. The cosmology induced on the brane recovers the standard form for the late universe. The higher-order correction becomes significant only in the early universe. In the following, we only focus on the GW events in the late universe, so the scale factor will follow the standard form for a vanishing spatial curvature density:
\begin{equation}
	H^{2}_{~}=H^{2}_{B}\Big(\Omega_{\Lambda_{\text{eff}}}+\frac{\Omega_{m}}{a^{3}_{~}}\Big),
\end{equation}
where $H^{~}_{B}$ is the present Hubble parameter, and $\Omega_{\Lambda_{\text{eff}}}$ and $\Omega_{m}$ are the present density parameters for the dark energy and the nonrelativistic matter, respectively.

Now let us consider the projections of the trajectories of GWs and EMWs onto the 4-brane in the model of brane cosmology. The point is that in the landscape of a braneworld model, gravitons can propagate freely in the bulk while other particles in the Standard Model are confined on the 4-brane. Therefore, even if a GW and an EMW both travel at the speed of light from the same source, the projections of their trajectories on the 4-brane may not coincide. For the sake of simplification, we will focus on those gravitons and photons that propagate only in $R$ and $r$ directions. So for a graviton following the $D$-dimensional null geodesic with $\text{d}\theta=\text{d}\phi=\text{d}\theta_{1}^{~}=\dots=\text{d}\theta_{d-1}^{~}=0$, its path is governed by
\begin{equation}\label{gpr1}
	\text{d}s_{D}^{2}=-R^{2}_{~}\text{d}T^{2}_{~}+A\text{d}R^{2}_{~}+R^{2}_{~}\text{d}r^{2}_{~}=0.
\end{equation}
On the three-dimensional submanifold ($R\times M_{2}^{~}$) upon which the path resides, we can further give two Killing vectors, $K^{M}_{T}=(1,0,0,\dots,0)$ and $K^{M}_{r}=(0,1,0,\dots,0)$. Thus the comoving observer $U^{M}_{~}=\text{d}x^{M}_{~}/\text{d}\lambda$ of this path will find two conserved quantities, $\kappa_{T}^{~}$ and $\kappa_{r}^{~}$, as follows:
\begin{subequations}
	\begin{eqnarray}
		\kappa_{T}^{~}
		\!&\!=\!&\!
		\tilde{h}_{MN}^{(3)}U^{M}_{~}K^{M}_{T}=-R^{2}_{~}\frac{\text{d}T}{\text{d}\lambda},\\
		\kappa_{r}^{~}
		\!&\!=\!&\!
		\tilde{h}_{MN}^{(3)}U^{M}_{~}K^{M}_{r}=R^{2}_{~}\frac{\text{d}r}{\text{d}\lambda},
	\end{eqnarray}
\end{subequations}
where $\tilde{h}_{MN}^{(3)}$ is the projection tensor of the subspace and $\lambda$ is the affine parameter of the geodesics. Substituting these conserved quantities into \eqref{gpr1}, we can obtain the following equations describing the path of the graviton:
\begin{subequations}\label{eog1}
	\begin{eqnarray}
		\Big(\frac{\text{d}R}{\text{d}\lambda}\Big)^{2}_{~}
		\!&\!=\!&\!
		\frac{-2\Lambda}{(d+2)(d+3)}(\kappa^{2}_{T}-\kappa^{2}_{r}),\label{eogR1}\\
		\Big(\frac{\text{d}T}{\text{d}\lambda}\Big)^{2}_{~}
		\!&\!=\!&\!
		\frac{\kappa^{2}_{T}}{R^{4}_{~}},\label{eogT1}\\
		\Big(\frac{\text{d}r}{\text{d}\lambda}\Big)^{2}_{~}
		\!&\!=\!&\!
		\frac{\kappa^{2}_{r}}{R^{4}_{~}}.\label{eogr1}
	\end{eqnarray}
\end{subequations}
Assuming that the graviton originates from point $A$ on the 4-brane, escapes into the bulk, and eventually returns to the 4-brane at point $B$, its path in the bulk can be projected along the $r$ direction to yield an effective distance through
\begin{equation}
	\text{d}r^{2}_{~}=\frac{(d+2)(d+3)}{-2\Lambda}\frac{1}{R^{4}_{~}}\frac{1}{s-1}\text{d}R^{2}_{~},
\end{equation}
where $s\equiv\kappa^{2}_{T}/\kappa^{2}_{r}$ and we have used Eqs.~\eqref{eogR1} and~\eqref{eogr1}. For a four-dimensional observer on the 4-brane, such an effective distance defines the gravitational horizon radius,
\begin{equation}\label{ghr1}
	r_{g}^{~}\equiv\int_{r_{A}^{~}}^{r_{B}^{~}}\text{d}r=\int_{R_{A}^{~}}^{R_{B}^{~}}\sqrt{\frac{(d+2)(d+3)}{-2\Lambda(s-1)}}\frac{1}{R^{2}_{~}}\text{d}R.
\end{equation}
This expression is not practicable yet, because the four-dimensional observer can never measure the values of $R_{A}^{~}$ and $R_{B}^{~}$ directly on the 4-brane. Recalling Eqs.~\eqref{eogR1} and~\eqref{eogT1}, one can obtain the following relation:
\begin{equation}\label{RTr1}
	\text{d}R^{2}_{~}=\frac{-2\Lambda}{(d+2)(d+3)}\frac{s-1}{s}R^{4}_{~}\text{d}T^{2}_{~}.
\end{equation}
It could help us to re-express the gravitational horizon radius~\eqref{ghr1} in terms of the bulk time interval between the two points:
\begin{equation}\label{ghr2}
	r_{g}^{~}=\frac{T_{AB}^{~}}{\sqrt{s}}.
\end{equation}
Since the bulk time is related to the cosmic time through the relation~\eqref{cbr1}, we can convert the bulk time interval into
\begin{equation}
	T_{AB}^{~}=\int_{a_{A}^{~}}^{a_{B}^{~}}\sqrt{1-\frac{(d+2)(d+3)}{-2\Lambda}H^{2}_{~}}\frac{1}{Ha^{2}}\text{d}a,
\end{equation}
where we have used the relation~\eqref{sfR1}. So far, the only unobservable quantity in the gravitational horizon radius~\eqref{ghr2} is the parameter $s$. To replace it with an observable quantity, we can use integral Eq.~\eqref{RTr1}. With Eq.~\eqref{sfR1}, we get
\begin{equation}
	T_{AB}^{2}=\frac{(d+2)(d+3)}{-2\Lambda}\frac{s}{s-1}\Big(\int_{a_{A}^{~}}^{a_{B}^{~}}\frac{1}{a^{2}_{~}}\text{d}a\Big)^{2}_{~}.
\end{equation}
So under the low-energy limit ($\frac{(d+2)(d+3)}{-2\Lambda}H^{2}_{~}\ll1$), the gravitational horizon radius can be analytically expressed in terms of the source redshift $z_{A}^{~}$ as
\begin{eqnarray}\label{ghr3}
	r_{g}^{2}
	\!&\!=\!&\!
	T_{AB}^{2}-\frac{(d+2)(d+3)}{-2\Lambda}\Big(\int_{a_{A}^{~}}^{a_{B}^{~}}\frac{1}{a^{2}_{~}}\text{d}a\Big)^{2}_{~}\nonumber\\
	\!&\!\approx\!&\!
	\frac{1}{H_{B}^{2}\Omega_{\Lambda_{\text{eff}}}}\Theta_{1}^{2}+\frac{(d+2)(d+3)}{-2\Lambda}\Theta_{1}^{~}\Theta_{2}^{~}-\frac{(d+2)(d+3)}{-2\Lambda}z_{A}^{2},
\end{eqnarray}
where we have expanded the bulk time interval up to the order $\sim H^{2}_{~}/\Lambda$. The variables $\Theta_{1}^{~}$ and $\Theta_{2}^{~}$ are defined by
\begin{subequations}
	\begin{eqnarray}
    	\Theta_{1}^{~}
    	\!&\!=\!&\!
    	_{2}F_{1}\Big(\frac{1}{3},\frac{1}{2};\frac{4}{3};-\frac{\Omega_{m}}{\Omega_{\Lambda_{\text{eff}}}}\Big)-(1+z_{A}^{~})_{\,2}F_{1}\Big[\frac{1}{3},\frac{1}{2};\frac{4}{3};-\frac{\Omega_{m}}{\Omega_{\Lambda_{\text{eff}}}}(1+z_{A}^{~})^{3}_{~}\Big],\\
    		\Theta_{2}^{~}
    	\!&\!=\!&\!
    	_{2}F_{1}\Big(\!-\frac{1}{2},\frac{1}{3};\frac{4}{3};-\frac{\Omega_{m}}{\Omega_{\Lambda_{\text{eff}}}}\Big)-(1+z_{A}^{~})_{\,2}F_{1}\Big[\!-\frac{1}{2},\frac{1}{3};\frac{4}{3};-\frac{\Omega_{m}}{\Omega_{\Lambda_{\text{eff}}}}(1+z_{A}^{~})^{3}_{~}\Big],
	\end{eqnarray}
\end{subequations}
where $_{2}F_{1}$ is the hypergeometric function. Note that, for the sake of simplicity, we have made the assumption that the redshift of point $B$ is zero. Consequently, the value of the scale factor at point $B$ can be set as $a_{B}^{~}=1$.

Unlike gravitons, the Standard Model particles are always confined on the 4-brane. For a photon, its trajectory follows the four-dimensional null geodesic on the 4-brane. Similar to the case of the graviton, we fix the start point of the four-dimensional null geodesic at point $A$, while setting the end point at $C$ without loss of generality. With the induced metric~\eqref{im1} and the relations \eqref{cbr1} and~\eqref{sfR1}, the four-dimensional null geodesic satisfies
\begin{equation}\label{4Dg1}
    -\text{d}t^{2}_{~}+a^{2}_{~}\text{d}r^{2}_{~}=0.
\end{equation}
Thus, the photon horizon radius is given by
\begin{equation}\label{phr2}
	r_{\gamma}^{~}=\int_{r_{A}^{~}}^{r_{C}^{~}}\text{d}r=\int_{t_{A}^{~}}^{t_{B}^{~}}\frac{1}{a}\text{d}t,
\end{equation}
where $r_{C}^{~}$ is the radial coordinate distance of point $C$. Here, we assume that the photon reaches point $C$ at the moment the graviton reaches point $B$, so we have $t_{C}^{~}=t_{B}^{~}$. For the model with a curved 4-brane, the projection of a higher-dimensional null geodesic onto the 4-brane usually deviates from any four-dimensional null geodesics localized on the 4-brane. So point $C$ does not need to overlap with point $B$. It finally results in a difference between the gravitational horizon radius and the photon horizon radius. And the four-dimensional observer will observe a time delay between the arrival of the graviton and photon.

To compare the photon horizon radius and the gravitational horizon radius, we convert the above expression~\eqref{phr2} into
\begin{equation}\label{phr1}
	r_{\gamma}^{2}=\left(\int_{a_{A}^{~}}^{a_{B}^{~}}\frac{1}{a^{2}_{~}H}\text{d}a\right)^{2}_{~}=\frac{1}{H_{B}^{2}\Omega_{\Lambda_{\text{eff}}}}\Theta_{1}^{2}.
\end{equation}
Obviously, the photon horizon radius happens to be the leading-order term of the gravitational horizon radius~\eqref{ghr3}. Since the rest terms of the gravitational horizon radius are positive definite, we have $r_{g}^{~}>r_{\gamma}^{~}$. So the $D$-dimensional null geodesic is always shorter than its counterpart in the four-dimensional space in our model. In other words, the graviton takes a shortcut. Consequently, a GW signal will always reach a four-dimensional observer earlier than its EM counterpart signal that is simultaneously triggered by the same source. If the expansion of the universe is negligible during the period $\Delta t$, we can arrive at an approximation on the time delay,
\begin{equation}\label{td1}
	c\,\Delta t\approx r_{g}^{~}-r_{\gamma}^{~},
\end{equation}
which may reveal the structure and number of extra dimensions.

\section{Limits}\label{sec4}

In the last section, we show an approach to investigate the structure and more importantly the number of extra dimensions by the shortcut effect in the joint observation of GWs and EMWs. However, it requires that the target signals are triggered by the same source, and that there is a predictable time interval between their emissions. In astrophysics, one of the target sources that satisfies these requirements is the merger of a BNS~\cite{Shibata1,Rezzolla1,Tsang1,Paschalidis1,Ciolfi1,Rezzolla2}. It was in 2017 that the joint GW and EMW observations found a GW event (GW170817) and a short gamma-ray burst event (GRB 170817A) emitted from the same source located at NGC 4993~\cite{Abbott6,Coulter1,Pan1,Abbott7}. In the event GW170817/GRB 170817A, the source is probed to be the coalescence of a BNS~\cite{Abbott7}, and the EMW signal arrived $\sim1.74\,\text{s}$ later than the GW signal. In the following, we will suppose that two signals in the event GW170817/GRB 170817A are triggered by the BNS simultaneously. We will also ignore the contribution from the intergalactic medium dispersion on the wave propagation for the sake of simplification. Under these assumptions, the time interval between the GW signal and the EMW signal is the only result from the existence of the shortcut. Note that the source redshift of the event GW170817 is only at the order of $\sim 0.01$~\cite{Abbott6}. It points to the late stages of the universe that is well described by the cosmological model that we use to derive Eqs.~\eqref{ghr3},~\eqref{phr1}, and~\eqref{td1}. Thus the bulk cosmological constant, the number of extra dimensions, and the observables in the event GW170817/GRB 170817A satisfy the following relation:
\begin{equation}\label{dt1}
	\Delta t^{2}\approx \frac{(d+2)(d+3)}{-2\Lambda}(\Theta_{1}^{~}\Theta_{2}^{~}-z_{A}^{2})+ \frac{2\Delta t\,\Theta_{1}^{~}}{H_{B}^{~}\sqrt{\Omega_{\Lambda_{\text{eff}}}}}.
\end{equation}
Here, we have $z_{A}^{~}=0.008^{+0.002}_{-0.003}$ and $\Delta t=1.74^{+0.05}_{-0.05}\,\text{s}$ according to the event GW170817/GRB 170817A. The 2018 release of Planck satellite data~\cite{Planck1} indicates $H_{B}^{~}\sim67.66\,\text{km}\,\text{s}^{-1}_{~}\text{Mpc}^{-1}_{~}$, $\Omega_{\Lambda_{\text{eff}}}\sim0.6889$, and $\Omega_{m}\sim0.3111$.

\begin{figure}[!htb]
\center{
\subfigure[]{\includegraphics[width=6cm]{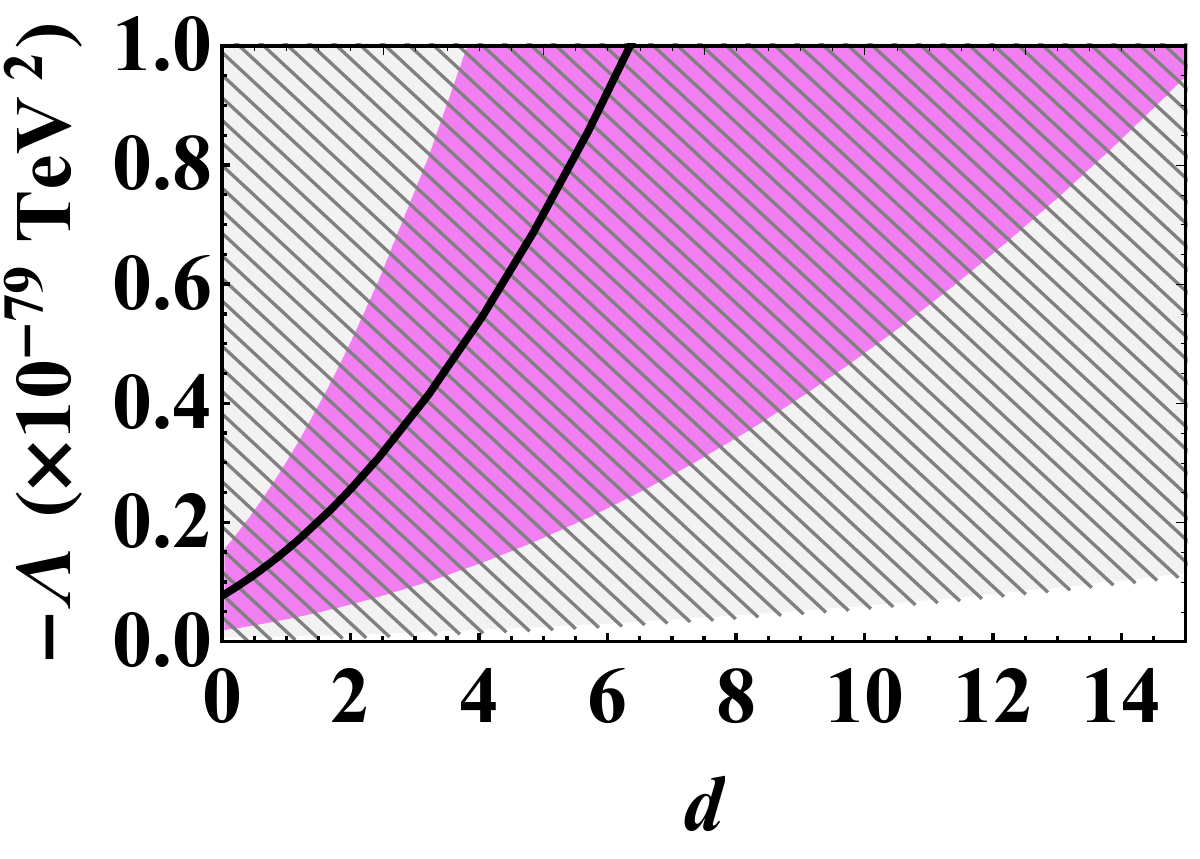}\label{Ld1}}
\quad\quad
\subfigure[]{\includegraphics[width=6cm]{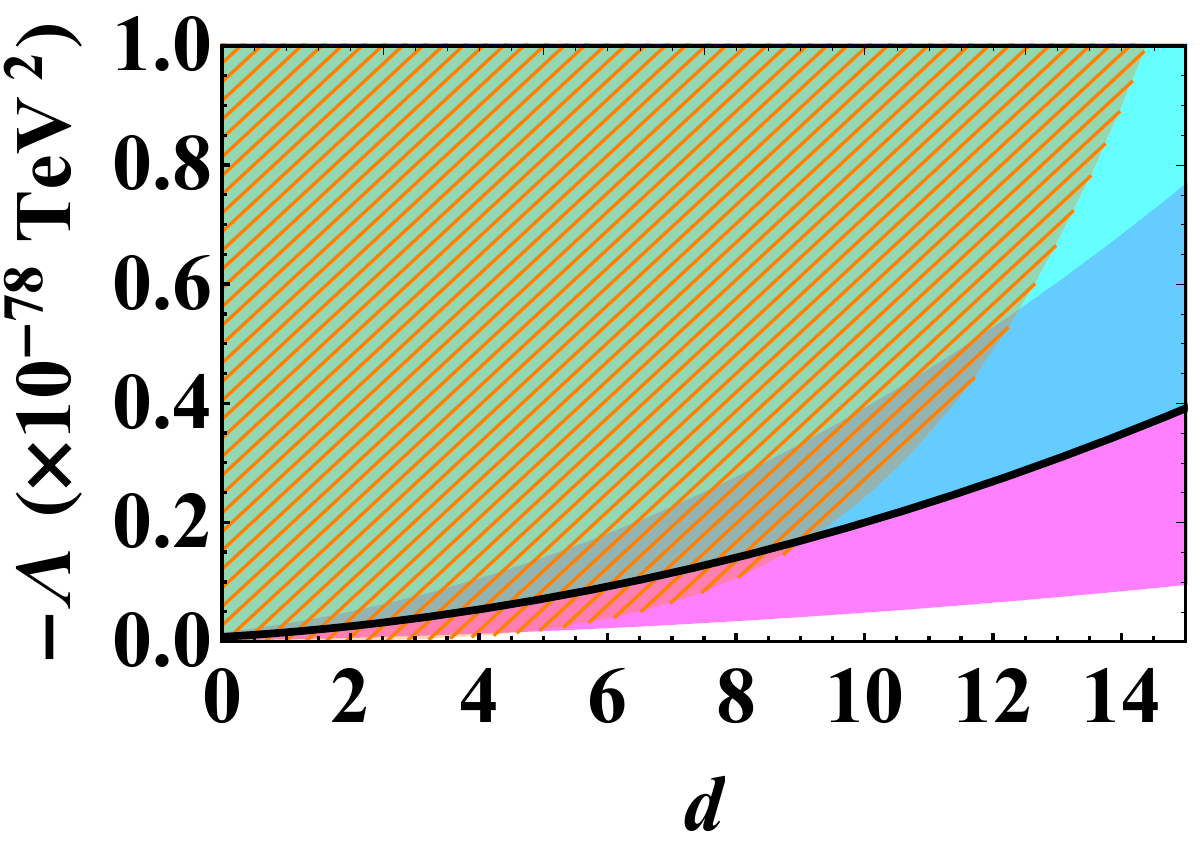}\label{Ld2}}
}
\caption{Constraints on the bulk cosmological constant and the number of extra dimensions. The black solid line corresponds to the case of $z_{A}^{~}=0.008$ and $\Delta t=1.74\,\text{s}$. The magenta area counts the errors on $z_{A}^{~}$ and $\Delta t$. (a) The gray meshed region is given by Eq.~\eqref{ecm1} with $VM^{D-2}_{*}\gtrsim10^{45}_{~}\,\text{TeV}^{3}_{~}$ given by Refs.~\cite{Visinelli1,Yu1}. (b) The cyan area counts the contributions from different astrophysical processes. It carries a window of the effective time delay as $(0\,\text{s}, 1.74\,\text{s})$. The orange meshed area is given by the model that considers the backreaction of a $\text{dS}_{4}^{~}$ brane to the bulk spacetime~\cite{Lin3}.}
\label{constraint1}
\end{figure}

In Fig.~\ref{constraint1}, we show the constraint on the bulk cosmological constant and the number of extra dimensions based on the event GW170817/GRB 170817A. If we ignore the the errors of the source redshift and time delay (see the black solid line), we have
\begin{equation}\label{adsr1}
	\frac{-2\Lambda}{(d+2)(d+3)}\sim10^{-81}_{~}\,\text{TeV}^{2}_{~}.
\end{equation}
On the other hand, according to $\Omega_{\Lambda_{\text{eff}}}$, the effective cosmological constant is $\Lambda_{\text{eff}}\sim10^{-90}\,\text{TeV}^{2}_{~}$. So it is much smaller than the bulk cosmological constant, satisfying the low-energy approximation in the paper. Recalling the relation~\eqref{ecm1}, we can directly estimate and obtain
\begin{equation}\label{es1}
	\frac{VM_{*}^{D-2}}{M^{2}_{\text{Pl}}}\sim10^{-41}_{~}\,\text{TeV}.
\end{equation}
Reminding that the volume $V$ arises from the Kaluza-Klein reduction of the constant-curvature spacelike dimensions ($S_{d-1}^{~}$), so this estimation~\eqref{es1} indeed places a constraint on the effective mass scale $\tilde{M}_{*}^{(5),3}\equiv VM^{D-2}_{*}$ of the effective five-dimensional theory. One can check that the estimated $\tilde{M}_{*}^{(5)}$ is consistent with the constraint ($\tilde{M}_{*}^{(5)}\gtrsim10^{-3}_{~}\,\text{TeV}$) given by Refs.~\cite{Visinelli1,Yu1}. Thus the embedding of a $\text{dS}_{4}^{~}$ brane in our model satisfies both the GW and cosmological observations. This result remains robust when considering the observational errors of the redshift and time delay because our model allows for a wide range of parameters for $\Lambda$ and $d$ (see Fig.~\ref{Ld1}). In fact, the stability requirement to the model can compress this region.

In Ref.~\cite{Lin3}, the moving 4-brane is treated as a small perturbation to the bulk spacetime. Thus, the backreaction of the 4-brane to the bulk spacetime is taken into account, which leads to corrections to the background solutions. With this perturbative analysis, the model is proved to be stable under the embedding of a nonlinear dynamical brane. Therein, the standard Friedmann equations can also be recovered in the late universe. The embedding of a $\text{dS}_{4}^{~}$ brane in the bulk would require the bulk cosmological constant to satisfy
\begin{equation}\label{Ly1}
	\frac{-2\Lambda}{(d+3)(d+2)}
		=
		\bigg(\frac{M^{D-2}_{*}V}{M_{\text{pl}}^{2}}\bigg)^{2}_{~}\mathcal{Y}^{2}_{~},
\end{equation}
where $\mathcal{Y}$ is a parameter function related to $d$. As is shown in Fig.~\ref{Ld2}, it provides a constraint on the parameter space of $\Lambda$ and $d$. With the observations of GW170817 and GRB 170817A, the number of extra dimensions is limited to $d\leq9$. Note that the allowed region for $\Lambda$ and $d$ can be expanded by including the errors on the source redshift and time delay. As is shown by the overlap of the magenta and orange meshed areas in Fig.~\ref{Ld2}, these errors can significantly shift our limit to $d\leq4$ and $d\leq12$ by the lower and upper boundaries of the magenta area, respectively. The error range of the result is dominated by the source redshift, because the error of redshift in the event GW170817/GRB 170817A is at the same order as the observed value. Therefore, a more precise measurement on the redshift would enable us to impose a more stringent constraint.

Note that the above analyses are based on the assumption that the GW signal and the EMW signal in the event GW170817/GRB 170817A are triggered by the source simultaneously. In fact, in different astrophysical models, there could be a time lag between their emissions by considering the collapse time of the remnant and the energy dissipation process during the merger of the BNS. Thus the effective time delay between the two signals can be larger or smaller than the observed one. A general prediction suggests that the effective time delay can be corrected up to a few seconds or down to zero~\cite{Li:2016zjc}. As is shown by the cyan area in Fig.~\ref{Ld2}, these astrophysical processes can further relax the parameter space of $\Lambda$ and $d$. The lower boundary of the cyan area overlaps the black solid line, which represents no time lag between emissions of two signals. If we suppose that the EMW signal launched $1.74\,\text{s}$ later than the GW signal, the effective time delay becomes $0\,\text{s}$. However, it does not imply that $d=0$ because the number of extra dimensions and the bulk cosmological constant are both related to the time delay [see~Eq.~\eqref{dt1}]. As is shown in Eq.~\eqref{ghr3}, when $\Lambda$ goes infinite, the gravitational horizon radius approaches to the photon horizon radius. In this case, the effective time delay tends to $0\,\text{s}$ (which corresponds to the upper boundary of the cyan area) and the number of extra dimensions is unconstrained. This is reasonable, since an extremely large bulk cosmological constant means that the propagation of a GW along $R$ direction is negligible compared to its propagation in other directions, and that trajectories of GWs and EMWs, ~\eqref{gpr1} and~\eqref{4Dg1}, approximately overlap. There is no stricter limit can be provided by the introduction of astrophysical models. Therefore, we can place a conservative limit to the number of extra dimensions as $d\leq12$.

In fact, the time delay also gives a constraint on the $\text{AdS}_{D}^{~}$ radius as $\ell^{2}_{D}\lesssim0.04\,\text{Mpc}^{2}$ through Eq.~\eqref{adsr1}. This constraint is not competitive with the current terrestrial experiments, such as the torsion balance measurements. This large constraint means that the correction to the $1/r^{2}_{~}$ law for gravity on the 4-brane could be significant at $r\lesssim\ell_{D}^{~}\sim\text{Mpc}$. We therefore set the $\text{AdS}_{D}^{~}$ radius as $\ell^{2}_{D}\lesssim3.65\times10^{-54}\,\text{Mpc}^{2}$ compatible with the current torsion balance measurements~\cite{Tan:2016vwu}. Then the time delay contributed from the shortcut effect becomes $\Delta t\lesssim7.78\times10^{-52}\,\text{s}$, which is negligible compared to the time delay observed in the event GW170817/GRB 170817A. In this case, contributions from astrophysical models are dominant, and, as we have discussed above, the number of extra dimensions becomes unconstrained.

\section{Conclusion}\label{sec5}

The number of extra dimensions is one of the fundamental questions in braneworld theories. The detection of GWs has opened up a new avenue for constraining this number. The leakage of gravitons during the propagation of GWs, as observed in the event GW170817/GRB 170817A, has provided a lower bound on the number of extra dimensions~\cite{Pardo1,Abbott8}. In this paper, our interest is on placing an upper bound for this number with the shortcut, which might help to narrow down the parameter space of higher-dimensional theories and potentially rule out some of them.

We constructed a nonlinear dynamical braneworld model by embedding a 4-brane under specific conditions. For the sake of simplicity, we assumed that the 4-brane has no backreaction to the bulk spacetime. To examine whether the cosmology on the 4-brane can yield the standard one predicted in GR, we derived the Israel joint condition. By imposing the embedding condition, we found that the balance between the bulk cosmological constant and the bare cosmological constant ensures the equations governing the brane cosmology reduce to the standard Friedmann equations at the leading order. Thus Eq.~\eqref{fFe1} could describe the normal expansion of our late-time universe. In addition to it, the contributions from extra dimensions are all encapsulated in the higher-order corrections to Eq.~\eqref{fFe1}, and their effects were apparent in the early universe. Remarkably, the joint condition could also describe the dynamics of the 4-brane. It was shown that the nonlinear dynamics of the 4-brane is sourced by the matter on the 4-brane.

We then considered the path of a GW emitted from and eventually returning onto the 4-brane. The target signal we focused on is originated from a low-redshift source.  Thus we can use the standard $\Lambda$CDM model to describe the expansion of the 4-brane during the propagation of the signal. By deriving the gravitational horizon radius, we ensured that it is larger than the photon horizon radius on the 4-brane within the same time frame. It means that the path of gravitons connecting any two points on the 4-brane is shorter than the one of photons in our model, supporting the existence of shortcuts. This feature might support the time delay observed in the event GW170817/GRB 170817A.

With the modified balancing between the bulk and bare cosmological constants in Ref.~\cite{Lin3}, we finally placed a limit to the number of extra dimensions as $d\leq9$. This result is sensitive to the value of the source redshift. We found that the upper bound on the number of extra dimensions can be significantly shifted from 4 to 12 by taking the errors into account. We then imposed a conservative limit to this quantity, i.e., $d\leq12$. It is robust under the consideration of astrophysical models. We therefore concluded that higher-dimensional theories with less than 16 extra dimensions are not yet ruled out by our analyses.

This result was based on the assumption that $\ell^{2}_{D}\sim0.04\,\text{Mpc}^{2}$. With the joint constraint from the torsion balance measurements, the $\text{AdS}_{D}^{~}$ radius became $\ell^{2}_{D}\lesssim3.65\times10^{-54}\,\text{Mpc}^{2}$. It significantly depressed the contribution from the shortcut effect on the time delay. We found that under this stricter constraint, the observed time delay is dominated by the contribution from astrophysical models. In this case, the number of extra dimensions was unconstrained. In other words, theories with more than 16 extra dimensions are also acceptable. However, we should note that the above underlining theory of gravity is GR. An alternative way to avoid the strong constraint $\ell^{2}_{D}\lesssim3.65\times10^{-54}\,\text{Mpc}^{2}$ was to introduce a nonminimal coupling between the gravity and the bulk scalar field, in which case four-dimensional GR could be recovered on the 4-brane at a small length scale while leaving a correction at a large length scale~\cite{Dvali1}. We will leave this work for the future.

\section*{Acknowledgements}

We acknowledge useful discussions with Yu-Xiao Liu, Yi Zhong, and Yuan-Chuan Zou. This work is supported by the National Key Research and Development Program of China (Grant No.~2020YFC2201503) and the National Natural Science Foundation of China (Grants No.~12247142 and No.~12047564).

\end{document}